\documentclass[twocolumn,
             superscriptaddress, 
              longbibliography, 
               showkeys,apl, floatfix]{revtex4-1}

\usepackage{lipsum}

\usepackage{graphicx} 
\usepackage{bm}	
\usepackage{color}                     
\usepackage{xcolor}
\usepackage{epsfig}
\usepackage{amsmath} 
\usepackage{amssymb} 
\usepackage{longtable} 
\usepackage{float}
\usepackage{mathtools}
\usepackage{xparse}
\usepackage{hyperref}
\usepackage{times}
\usepackage[normalem]{ulem}

\newcommand{\Lun}{L}

\usepackage[ruled]{algorithm}
\usepackage{algpseudocode}
\usepackage{enumitem}

\usepackage{sidecap}
\sidecaptionvpos{figure}{t}

\newcommand{\BEA}{\begin{eqnarray}}
\newcommand{\EEA}{\end{eqnarray}}
\newcommand{\BEQ}{\begin{equation}}
\newcommand{\EEQ}{\end{equation}}
\newcommand{\BIT}{\begin{itemize}}
\newcommand{\EIT}{\end{itemize}}
\newcommand{\BNUM}{\begin{enumerate}}
\newcommand{\ENUM}{\end{enumerate}}

\newcommand{\BA}{\begin{array}}
\newcommand{\EA}{\end{array}}

\newcommand{\permutationinv}[1]{\pi^{-1}(#1)}

\newcommand{\be}{\begin{equation}}
\newcommand{\ee}{\end{equation}}
\newcommand{\bd}{\begin{displaymath}}
\newcommand{\ed}{\end{displaymath}}
\newcommand{\BE}{\begin{eqnarray}}
\newcommand{\EE}{\end{eqnarray}}

\newcommand{\R}{{\rm I\!R}}

\newcommand{\id}{{\rm 1\!\!I}}

\newcommand{\tr}{\mathrm{Tr}}

\newcommand{\KL}[2]{D_{\mathrm{KL}}(#1 || #2)}
\newcommand{\PBM}{P_{\mathrm{BM}}}
\newcommand{\NLL}{\mathrm{NLL}}
\newcommand{\corei}{\mathcal{A}^{(i)}}
\newcommand{\C}{\mathbb{C}}
\newcommand{\N}{\mathbb{N}}
\newcommand{\A}{\mathcal{A}}
\newcommand{\G}{\mathcal{G}}
\newcommand{\D}{\mathcal{D}}
\newcommand{\x}{\mathbf{x}}

\definecolor{darkgreen}{rgb}{0.0, 0.5, 0.0}

\begin{document}

\setlist[enumerate,1]{label=\arabic*, start=0}

% \apo{TO BE UPDATED}
\title{Qubit seriation: Improving data-model alignment using spectral ordering}

\author{Atithi Acharya}
\affiliation{Zapata Computing Canada Inc., 325 Front St W, Toronto, ON, M5V 2Y1}
\affiliation{Rutgers University, Piscataway, NJ 08854, USA }

\author{Manuel Rudolph}
\affiliation{Zapata Computing Canada Inc., 325 Front St W, Toronto, ON, M5V 2Y1}

\author{Jing Chen}
\affiliation{Zapata Computing Canada Inc., 325 Front St W, Toronto, ON, M5V 2Y1}

\author{Jacob Miller}
\affiliation{Zapata Computing Canada Inc., 325 Front St W, Toronto, ON, M5V 2Y1}

\author{Alejandro Perdomo-Ortiz}
\email{alejandro@zapatacomputing.com}
\affiliation{Zapata Computing Canada Inc., 325 Front St W, Toronto, ON, M5V 2Y1}

%end authors

\date{\today} 

\begin{abstract}

% \apo{Insert abstract here} 

\vspace{7pt}
With the advent of quantum and quantum-inspired machine learning, adapting the structure of learning models to match the structure of target datasets has been shown to be crucial for obtaining high performance. Probabilistic models based on tensor networks (TNs) are prime candidates to benefit from data-dependent design considerations, owing to their bias towards correlations which are local with respect to the topology of the model. In this work, we use methods from spectral graph theory to search for optimal permutations of model sites which are adapted to the structure of an input dataset. Our method uses pairwise mutual information estimates from the target dataset to ensure that strongly correlated bits are placed closer to each other relative to the model's topology. We demonstrate the effectiveness of such preprocessing for probabilistic modeling tasks, finding substantial improvements in the performance of generative models based on matrix product states (MPS) across a variety of datasets. We also show how spectral embedding, a dimensionality reduction technique from spectral graph theory, can be used to gain further insights into the structure of datasets of interest.
\vspace{3pt}
\end{abstract}

\maketitle

\vspace{4pt}

\section{Introduction}\label{s:intro}

The development of increasingly powerful quantum computers has placed renewed focus on the near-term potential of quantum models and algorithms for solving problems of significant real-world value. Probabilistic modeling, where a model is trained to learn the structure of an unknown distribution from an unlabeled dataset of samples, has emerged as an application of particular promise for quantum methods~\cite{PerdomoOrtiz2017}, owing to provable advantages in expressivity~\cite{gao2021enhancing} and generalization~\cite{gili2022evaluating} arising from the distinct properties of quantum state spaces. Within this domain, the use of quantum-inspired tensor networks (TNs) has allowed many of the advantages of fully-quantum probabilistic models to be enjoyed in a simulated classical setting~\cite{glasser2019, Bradley_2020}, while also permitting the development of hybrid quantum-classical models that exploit the complementary properties of both model families for practical benefit~\cite{gao2017efficient, Huggins_2019, wall2021generative, rudolph2022synergistic}.

Although promising, the relative newness of quantum and quantum-inspired machine learning algorithms means that best practices for ensuring optimal performance remain unsettled. While a large amount of attention has been dedicated to overcoming the phenomenon of \textit{barren plateaus} in optimization landscapes~\cite{Mcclear2018Barren, cerezo2021costfunction, holmes2022expressivity}, we focus here on a less well-understood issue, namely, the impact of the dataset's geometry on the performance of the machine-learning (ML) models.  
%While quantum entanglement forms a powerful modeling resource for reproducing complex correlations in data, it also remains a resource that is difficult to realize at long ranges, owing to limited gate connectivities in fully-quantum devices, and prohibitive classical computational overhead in quantum-inspired models. 
The importance of optimally matching model geometry to the structure of probabilistic modeling problems is well-understood in the TN community~\cite{evenbly2011tensor, lu2021tensor, convy2022mutual}. However, prior proposals for ensuring this optimal matching have tended to rely on heuristic search through various model configurations~\cite{li2020evolutionary, hashemizadeh2020adaptive, li2022permutation}, entailing a high cost due to repeated model retraining, while also failing to make use of insights present in the structure of the classical data itself. A notable exception is the problem-specific solution of Barcza et al.~\cite{Barcza_2011}, where \newline spectral ordering methods were shown to be capable of improving the performance of density matrix renormalization group (DMRG) calculations within quantum chemistry problems. Recent efforts to incorporate geometric considerations into quantum machine learning (QML), although of a different flavor than considered here, include the works of \cite{meyer2022exploiting,Larocca_2022,ragone2022representation,nguyen2022theory}, where the authors incorporate geometric priors arising from problem-specific symmetries into quantum models. \vspace{1.5pt}  \newline 
In this work, we introduce a simple method for ordering the variables of a classical dataset to ensure an optimal match between the correlations present in the data and the connectivity of 1D quantum or quantum-inspired models. We refer to this problem as \textit{qubit seriation}, in recognition of its similarity with the seriation problem of linearly-ordering sequential data, which has proven important in domains as diverse as archaeology~\cite{robinson_1951}, DNA sequencing~\cite{convex_rel, Atkins_1998}, and natural language processing~\cite{Order_matters}. Our approach makes use of tools from  spectral graph theory to efficiently compute an ordering directly from the pairwise mutual information between variables in a classical dataset of interest, thus guaranteeing that strongly correlated variables are mapped to nearby qubits, and weakly correlated variables to more distant qubits. We demonstrate the effectiveness of our procedure in probabilistic modeling experiments utilizing matrix product states (MPS), where reordering the variables of a classical dataset prior to optimization is shown to significantly boost the performance of the trained model. We show how spectral embedding tools can be used to extend these methods to models with more complex connectivities, and develop heuristics for understanding the impact of noise or small dataset size on the output ordering. Overall, our work emphasizes the practical importance of geometric considerations in quantum and quantum inspired machine learning, and demonstrates the performance benefits that are possible with the use of principled approaches to solving these issues. 

\section{Background}\label{s:background}

\subsection{Spectral Graph Theory and Mutual Information}

% \jem{I'm copying some material here from later in the paper, and we can make these snippets into something more coherent soon.}

The central object within spectral graph theory is graph Laplacian matrices. Although there exists several varieties of graph Laplacians, this work will focus on those based on undirected weighted graphs, which are described using a symmetric \textit{weight matrix} $W \in \R^{n \times n}$, whose nonnegative entries $W:=(w_{ij})$ specify the edge weights between the $n$ nodes of the corresponding graph. The (unnormalized) graph Laplacian is then given by $L = D - W$, where $D$ is the diagonal \textit{degree matrix} with entries of $D_{ij} = \delta_{ij} \sum_{j} W_{ij}$. 
% \mr{I feel like you should already mention that the graphs represents the pair-wise mutual information. :: Jacob and I believe its better to keep it for later as it is not crucial to spectral graph theory itself} 
The eigenvalues and eigenvectors of the graph Laplacian can be used to describe and study many properties of their respective graphs~\cite{Ulrike, Mohar1997}. Some important properties of the graph Laplacian include \cite{Ulrike}: (1) For every vector $f \in \R^n$ we have 
\begin{align}
\label{eq:laplacian_quadratic_identity}
f^T \Lun f = \frac{1}{2} \sum_{i,j=1}^n w_{ij} (f_i - f_j)^2.
\end{align}
(2) $\Lun$ is symmetric and positive semi-definite. (3) $\Lun$ has $n$ real, non-negative eigenvalues $0 =\lambda_0 \leq \lambda_2 \leq \hdots \leq \lambda_{n-1}$, with the (trivial) eigenvector associated with $\lambda_0$ being the all-ones vector $\x_{0} = \mathbf{1}$. (4) The number of zero eigenvalues $\lambda_i = 0$ is equal to the number of connected components in the graph, such that a graph with $k$ connected components will have $\lambda_k$ be the first non-zero eigenvector. In the following we will typically assume the use of connected graphs, so that $\lambda_1 > 0$.

The eigenvector associated with the first non-zero eigenvalue is referred to as the \textit{Fiedler vector}, and has a central role in many applications stemming from spectral graph theory. One important application lies in spectral clustering~\cite{Spectral_clustering} which uses spectral properties of the similarity graph Laplacians. Given a dataset, the goal is to build a similarity graph which models the local neighborhood relationships between the data points. These similarity graphs can be of various kinds such as $\epsilon$-neighborhood graph, $k$-nearest neighbor graphs, but our work will focus on fully connected graphs. Once a similarity graph is established, the graph Laplacian $L$ can be computed from the corresponding weighted weight matrix $W$. Then we can directly use the first m eigenvectors stacked as columns in a matrix $U\in R^{n \times m}$. Performing k-means clustering in this lower-dimensional subspace leads to an effective clustering of the original data points. Beyond spectral clustering, we will utilize the lesser-known formalism of spectral graph ordering~\ref{sec:spec_order}, which will be discussed in Sec.~\ref{sec:spec_order}.

Our work utilizes a similarity graph based on pairwise mutual information, which is simply the mutual information (MI) between each pair of random variables in a distribution. The MI between variables $X_i$ and $X_j$ is defined as $I(X_{i};X_{j}) = \KL{P_{X_{i}, X_{j}}}{P_{X_{i}}P_{X_{j}} }$, where $\KL{P}{Q} = \sum_x P(x) \left(\log(P(x)) - \log(Q(x))\right)$ denotes the Kullback-Leibler (KL) divergence.
% \mr{Notation not consistent with the NLL later on.}
Pairwise MI is not a distance metric between variables in a distribution, and in cases that a proper metric is needed, one can use (for example) a variation of information.
The sample complexity for estimating MI is well-studied in the information theory literature~\cite{Paninski}, and in the following we employ a maximum likelihood approach to estimate the MI between pairs of variables based on a number of samples from the underlying statistical distribution.

%\jem{Need a quick summary of definitions involving weighted graphs and graph Laplacians, along with some key properties of these Laplacians which are relevant to the following (e.g. lowest eigenvalue is zero and this is unique iff the graph is connected). There might be some other spectral graph theory definitions which are needed here.}

\subsection{Tensor Networks and Born Machines}

Tensor networks (TNs) are a family of models for describing large tensors $\psi \in \C^{d_1 \times d_2 \times \cdots \times d_n}$ using smaller tensor ``cores'', which are contracted together in a manner described by a defining graph $\G$~\cite{orus2014practical}. We focus on the case of matrix product states (MPS), whose $n$ cores $\{ \corei \in \C^{\chi_{i-1} \times d_i \times \chi_i} \}_{i=1}^n$ are contracted on a line graph along ``bonds'' of dimension $\chi_i \in \N$, whose size determines the expressivity of the model~\cite{garcia_mps}. Concretely, this describes a tensor with elements given by
\begin{equation}
    \psi_{x_1, x_2, \ldots, x_n} = \A^{(1)}_{x_1} \A^{(2)}_{x_2} \cdots \A^{(n)}_{x_n} ,
\end{equation}
where $\A^{(i)}_{x_i} \in \C^{\chi_{i-1} \times \chi_{i}}$ denotes the matrix with elements $(\A^{(i)}_{x_i})_{\alpha, \beta} = \A^{(i)}_{\alpha, x_i, \beta}$ and the indices associated with trivial bond dimensions $\chi_0 = \chi_n = 1$ are taken to be $x_0 = x_n = 1$. In the common setting where $\psi$ describes an $n$-body wavefunction, $\psi$ must be normalized to have unit Frobenius norm, i.e. $\lVert \psi \rVert_2^2 = \sum_{x_1, \ldots, x_n} |\psi_{x_1, \ldots, x_n}|^2 = \sum_{\x} |\psi_{\x}|^2 = 1$, where we use $\x = (x_1, \ldots, x_n)$ to denote the collection of all $n$ discrete indices of an MPS, which in our setting will describe the possible values of $n$ discrete variables. %\mr{this paragraph seems a little abstract to me :D }

Inspired by their long-established use in simulating many-body quantum systems, MPS have more recently been adapted to the task of learning classical probability distributions~\cite{ferris2012perfect, han2018unsupervised}, where they are referred to as MPS \textit{Born machines} (BMs). Each BM over an $n$-core MPS defines a probability distribution over $n$ discrete random variables, as $\PBM(\x) = |\psi_{\x}|^2$, which is properly normalized iff $\lVert \psi \rVert_2^2 = 1$. The $n$ cores of the MPS are then optimized to minimize a loss function measuring the compatibility of $\PBM$ with an unlabeled dataset $\D = \{ \x_t \}_{t = 1}^{T}$, which is typically chosen as the negative log likelihood loss $\NLL(\PBM, \D)$, defined by
\begin{align}
    \NLL(\PBM, \D) &= -\frac{1}{T} \sum_{t = 1}^{T} \ln(\PBM(\x_t)) \\
                   &= \KL{P_{\D}}{\PBM} + \ln(T),
\end{align}
where $P_{\D}$ is the empirical distribution given by $P_{\D}(\x) = 1 / T$ for $\x \in \D$ and $P_{\D}(\x) = 0$ otherwise. Given that $\NLL(\PBM, \D)$ is equal to $\KL{P_{\D}}{\PBM}$ up to a constant offset, minimizing the former is equivalent to minimizing the latter, which encourages $\PBM$ to assign large probability to samples $\x_t$ contained in $\D$.

Although MPS are ubiquitous in applications of TNs to quantum simulation, machine learning, and other fields, their 1D connectivity leads to a preference for capturing structure associated with sites $i, j$ for which $|i - j|$ is small. The primary reason for this preference is the limited capacity of MPS BMs, where the value of each bond dimension $\chi_i$ places an upper bound on the achievable MI between random variables associated with sites separated by the corresponding bond. Long-range correlations between two sites must be propagated through all intermediate bonds, which leads to a greater saturation of the capacity of the model than for short-range correlations. At an empirical level, it leads MPS BMs to learning target distributions more rapidly, and with smaller model sizes, when strongly correlated random variables are in close proximity to each other relative to the line graph defining the MPS.

\begin{figure}[t]
    \centering    
    \includegraphics[width=1.0\linewidth]{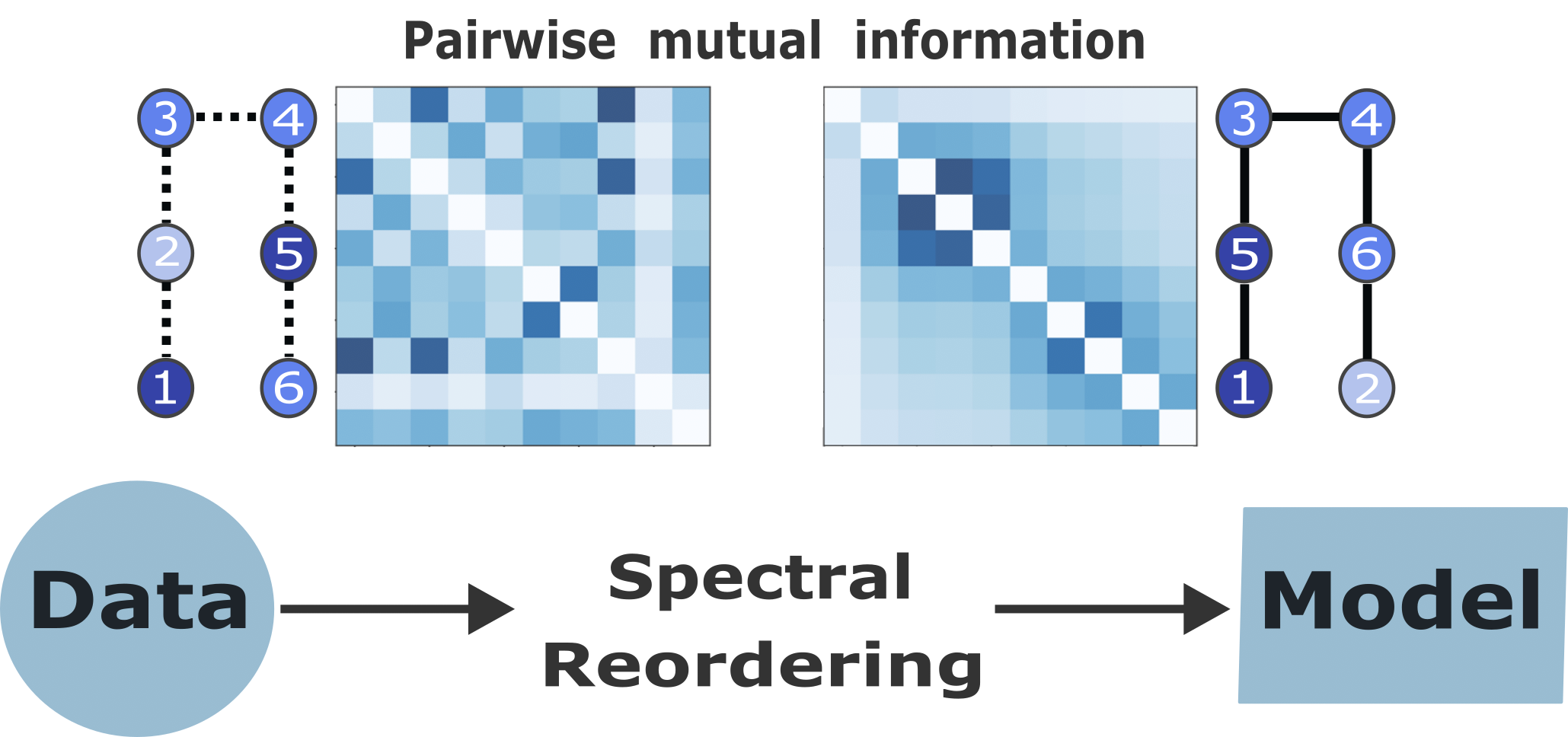}
    \caption{% Spectrally ordered data that restores the local structures is then fed into the training model.
    Schematic representation of our qubit seriation framework: Training data is often not structured such that neighboring sites of data samples are maximally correlated. However our spectral reordering preserves locality, as seen in the pairwise mutual information plots. We can see the unstructured data to the left has correlated qubits farther from each other but after we employ spectral ordering, the pairwise mutual information matrix has larger elements closer to the diagonal, as seen to the right of the figure. This reordering improves the performance of machine learning algorithms, such as those utilizing tensor network models.} 
    \label{fig:problem}
\end{figure}

\section{Methods}
\subsection*{Qubit seriation using spectral graph ordering}
\label{sec:spec_order}

% This figure belongs in later sections, but was moved here to encourage earlier placement in the paper
\begin{figure*}[ht]
    \centering
    \includegraphics[width=17.9cm]{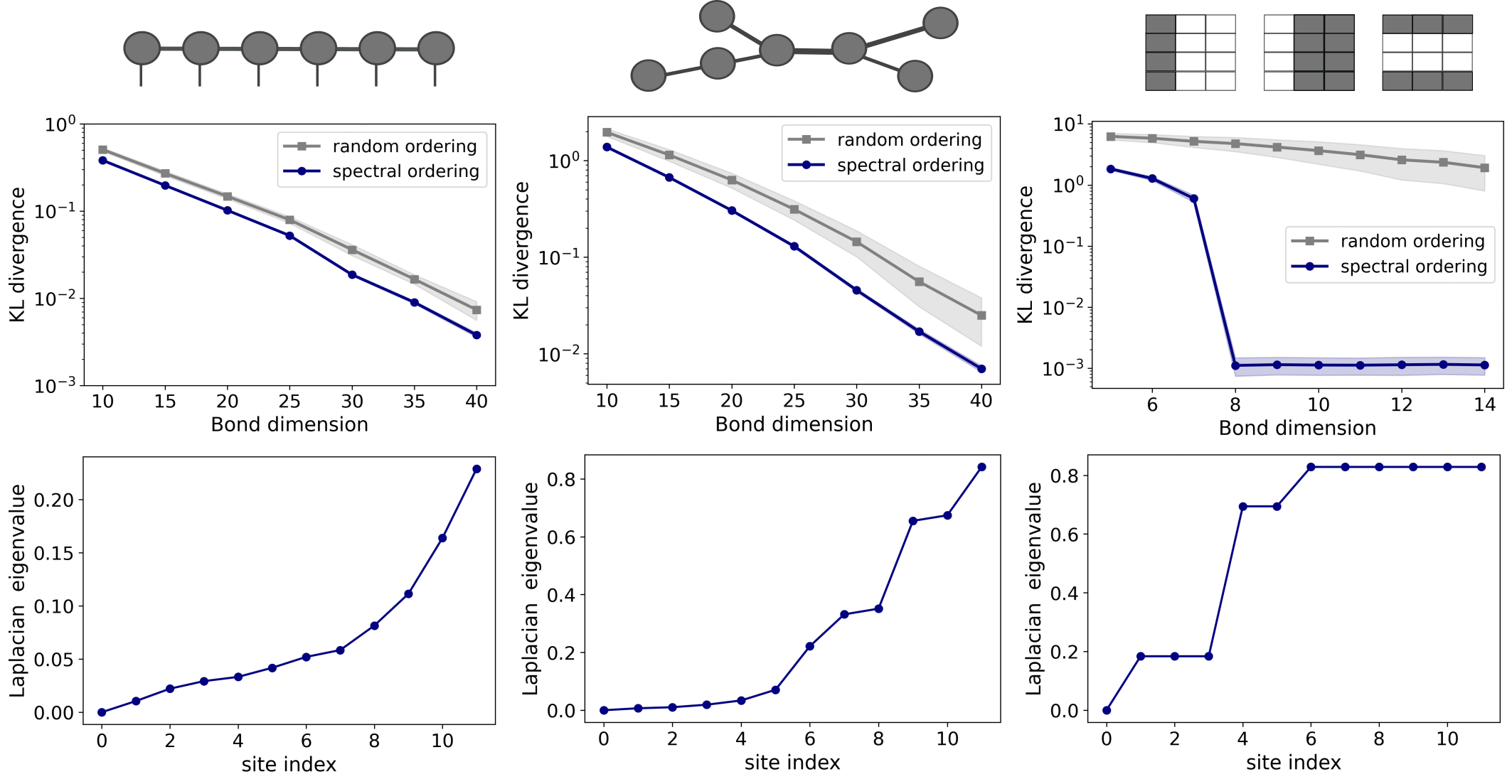}
    \caption{We plot the final KL divergence between the several training data distributions and trained MPS model distribution using random or seriated feature ordering. We have data generated by a random MPS to the left, followed by samples from a 12-qubit Ising tree Hamiltonian in the center, and finally the $4 \times 3$ bars and stripes dataset to the right. The performance of MPS trained using data with random site ordering are marked with gray, while the corresponding results using data with spectral ordering are marked with blue. The gray curves depict the median over 1000 different shufflings of the original dataset, while the fluctuations over these runs are indicated by shaded regions around the plots. In all three cases, the spectral ordering solution leads to a lower training error with greater than ($99 \%$) confidence. In the bottom row, we plot the eigenvalues of the corresponding Laplacian.}
    \label{fig:result}
\end{figure*}

We begin with an outline of a data-driven means of sequentially ordering variables within a distribution, which ensures that more strongly correlated sites are kept closer together than more distant sites. At a high level, this method employs a simple cost function scoring variable permutations based on the extent of long-range correlations in the reordered data. Although optimizing this cost function exactly is likely infeasible, we adopt a spectral ordering solution which uses spectral graph theory~\cite{Ulrike} to solve a convex relaxation of this original problem. We then develop a concrete connection between qubit seriation and spectral ordering, which makes use of the graph Laplacian associated with pairwise MI statistics in the data.

For data with $n$ sites, the ordering can be defined as the index permutation $\pi(1,2,\ldots,n)=(\pi(1), \pi(2), \ldots, \pi(n) )$. 
% Here we assume a system of n qubits. 
Each choice of ordering $\pi$ acts on the weight matrix to give a new matrix $(\pi W\pi^{T})_{ij}=w_{\pi(i),\pi(j)}$ expressing the MI between the permuted variables, and we aim to find a choice of $\pi$ such that larger values $w_{i,j}$ of the original weight matrix are mapped to values $w_{\pi(i),\pi(j)}$ such that $|\pi(i) - \pi(j)|$ is as small as possible, i.e., larger values in $W$ should be closer to the diagonal. One way of accomplishing this goal is to minimize a permutation-dependent cost function $C_{\mathrm{perm}}(\pi)$, which measures the extent of long-range correlations in the permuted dataset. A straightforward choice of this cost function, which we will see enables the application of useful previous results on spectral ordering, is
% \mr{Be more verbose. What is this cost function for?} 
% One choice of a cost function for distance sensitive ordering is
%
%%%% JACOB'S SUGGESTION
% \begin{align} \label{eqn:disc_cost}
% C_{\mathrm{perm}}(\pi) &= \frac{1}{2}\sum_{i,j}(i-j)^2 w_{\permutationinv{i},\permutationinv{j}} \nonumber \\
% &=\frac{1}{2}\sum_{i,j}(\permutation{i} - \permutation{j})^2 w_{i,j}.
% \end{align}
% \jem{I changed the above equation to emphasize the reordering of the actual sites to minimize the (MI-weighted) average distance squared between them. Let me know if you'd prefer a different way of doing this. Atithi: I don't think this change is needed. }
%%%% ATITHI'S SUGGESTION
\begin{align} \label{eqn:disc_cost}
C_{\mathrm{perm}}(\pi) &= \frac{1}{2}\sum_{i,j}(i-j)^2 w_{\pi(i),\pi(j) } 
\end{align}
While finding the minimum of this cost function is a combinatorial optimization problem which likely cannot be achieved with any polynomial-time algorithm~\cite{Linear_Ding}, finding a low-cost ordering is nonetheless possible by solving a convex relaxation of this problem. More precisely, the Fiedler vector solves a convex relaxation of $C_{\mathrm{perm}}$ in Eq. \ref{eqn:disc_cost}, which is phrased in terms of vectors $\x$ whose entries are continuous variables $x_{i}\in [-1,1]$. Additionally, the shifting necessary in the discrete variables introduces a constraint $\sum_{i}x_{i}=0$. The cost function for this convex relaxations is given by
\begin{align}
    C(\x) &= \frac{1}{2}\sum_{i,j}(x_i - x_j )^2 w_{i,j}
    %= \sum_{i}x^2_{i}d_{i}-\sum_{i,j}x_{i}x_{j}w_{i,j}\\ \x^{T}(D-W)\x 
    = \x^{T}L\x,
    \label{eq:cost_fun2}
\end{align}
where $L$ denotes the graph Laplacian associated to the matrix of pairwise MI values $w_{i,j}$, and the second equality comes from Eq.~\ref{eq:laplacian_quadratic_identity}. See Refs.~\cite{convex_rel,Atkins_1998} for a similar analysis in the context of DNA sequencing.

% \jem{Note that Eqs.~\ref{eq:cost_fun2} and \ref{eq:cost_fun2_unconstrained} aren't actually functions of a permutation $\pi$. I've rewritten them as $C(\x)$ instead (they were both written as $C(\pi)$ before), but let me know if something else would be better here. Atithi: I am fine with this change  }

% \jem{Also, I rewrote the cost functions in Eqs.~\ref{eqn:disc_cost}, \ref{eq:cost_fun2}, and \ref{eq:cost_fun2_unconstrained} (which had all been written as $C$) as $C_{\mathrm{perm}}$, $C$, and $C'$ respectively. Let me know if you would prefer something else here.}

To avoid trivial solutions, this convex relaxation requires the additional constraint $\x^T\x=1$, which can be imposed using a Lagrange multiplier $\lambda$.
% \mr{Maybe choose a different symbol for the lagrange multiplier, because of the eigenvalues. Actually they are the same: I can show you the maths} 
The stationary points of this cost function occur when $\x$ is an eigenvector of the positive semidefinite operator $L$. Now the cost function reads as 
\begin{equation}
    C'(\pi)=\x^{T}L\x-\lambda(\x^T\x-1),
    \label{eq:cost_fun2_unconstrained}
\end{equation} 
whose minimization is equivalent to minimizing the Rayleigh quotient $R(x)=\frac{\x^{T}L\x}{\x^T\x}$. It is straightforward to verify that the minimization of Eq.~\ref{eq:cost_fun2_unconstrained} subject to the constraint $\sum_{i}x_{i}=0$ is accomplished by the Fiedler vector $\x_1$. %We note that the lowest eigenvalue of the Laplacian is always $\lambda_{0}=0$ with the eigenvector being the trivial solution $\x_0=(1,1,...,1)^T$. The other $n-1$ eigenvectors are orthogonal to $\x_0$ which then satisfies the $\sum_{i}x_{i}=0$ condition naturally. This helps us in identifying that $\x_{1}$ minimizes the cost function while satisfying the constraints.

The solution of Eq.~\ref{eq:cost_fun2_unconstrained}, a convex relaxation of Eq.~\ref{eqn:disc_cost}, yields a vector $\x_1$, but our original goal was to identify an optimal permutation $\pi$. The link between these two problems arises by sorting the elements of $\x_{1}$ in ascending order, with the resulting permutation giving a heuristic solution to Eq.~\ref{eqn:disc_cost} that is provably optimal for certain families of weighted graphs (see Appendix~\ref{sec: stability_app} for more information). 
% By sorting the elements in $\x_{1}$ in ascending order and re-arranging the indices accordingly, we achieve the the optimal ordering for the original qubit seriation problem. 
Other ways of relaxing the cost function of Eq.~\ref{eqn:disc_cost} exist and can be exploited for some noisier datasets. One instance of the noise can be attributed to the estimation of the similarity values $w_{ij}$ from samples. 
% \jem{It would be helpful to have one sentence earlier in the paper which states exactly what is meant by ``noise'' in a dataset. Does this just mean small sample size (leading to statistical noise in the recovered MI matrix), or is something else implied here?} 
See Appendix~\ref{s:Proof_ordering} for a brief introduction to the relaxation using doubly stochastic matrices and the paper~\cite{convex_rel} for more details. 

%%%%%%%%% MI BASED %%%%%%%%%%%%%%%%% 
% \textbf{QUBIT SERIATION}:: 

% The pairwise MI statistics of the data can be understood as an undirected graph over the data sites with non-negative edges, as $I(X_{i}; X_{j})\geq 0$ is always positive by definition. These pairwise statistics are also used to characterize quantum properties in many-body systems \cite{Garcia_2020}.
% Here, we are trying to ensure that after the data has been ordered, the adjacent sites have high MI, i.e., they are more strongly correlated, whereas distant sites are less correlated. This can be seen in Fig.~\ref{fig:problem}, where, after ordering, the next neighbor sites tend to have large MI values. Thus we propose this method for qubit seriation.

% \jem{I commented out the final paragraph, since it is essentially restating things we've already said about pairwise MI and our use of this for qubit seriation. Atithi: OK}
%\input{sections/methods}
\section{Results}\label{s:results}

% \subsection{Qubit seriation using spectral graph reordering}

% \subsection{Similarity measure: Mutual Information}
% \label{sec:MI}

% The new methods introduced in our work are outlined and experimentally validated in 

Our spectral ordering method for qubit seriation is experimentally validated in three sections. 
In Sec.\ref{sec:spec_reorder} we verify the performance improvement of an MPS-based generative model trained on several datasets when using a spectral ordering of the underlying random variables. In Sec.~\ref{sec:spec_embedding} we then show how higher eigenvectors of the graph Laplacian can be used to reveal additional structural information present in the original dataset. 
Lastly, in Sec.~\ref{Sec:Stability} we analyze the impact of dataset size on the stability of the ordering arising from spectral ordering, which can be assessed using the spectral gap of the graph Laplacian.
%can be studied to indicate the success or failure of the spectral-based solutions. For substantiation, we monitor the spectral gaps in datasets consisting of a varied number of samples which leads to noisy estimates of similarity. 
\subsection{Spectral ordering based seriation in MPS based models}

\label{sec:spec_reorder}
We demonstrate the success of the solution to qubit seriation with 1D, 2D and other tree-structured data. While training the MPS based generative model using a negative log-likelihood (NLL) loss, we monitor the KL divergence $\KL{P_{\D}}{\PBM}$ between the training data distribution $P_{\D}$ and MPS distribution $\PBM$ throughout training.

% \begin{equation}\label{eq:KL_divergence}
% \begin{aligned}
%     \mathcal{L}\left(\theta\right) = \text{KL}\left(p_\mathcal{T} || q_\theta \right) & = \mathbb{E}_{\textbf{x}\sim p_\mathcal{T}(\textbf{x})}\left[\log \frac{p_\mathcal{T}(\textbf{x})}{q_\theta(\textbf{x})} \right]\\
%     & = - \log\left(|\mathcal{T}|\right) - \frac{1}{|\mathcal{T}|} \sum_{\textbf{x} \in \mathcal{T}} \log\left(q_\theta(\textbf{x})\right).
% \end{aligned}
% \end{equation}
% \mr{This has already been introduced in an earlier formula in the TN section.}
Fig.~\ref{fig:result} depicts our numerical simulations on random MPS, the ground states of random tree-structured Ising Hamiltonians, and the bars and stripes (BAS) dataset~\cite{han2018unsupervised, Benedetti2019}. Overall, it appears that qubit seriation leads to a better solution in more than $99\%$ of the total $1000$ experiments for all three chosen datasets. A final training loss is accepted to be better if it outperforms the randomly-ordered model's loss by more than a $1\%$ margin. 
% $\mathcal{L}_{random}-\mathcal{L}_{seriation}$ is greater than a $1\%$ margin where $\mathcal{L}$ is defined in Eq.\eqref{eq:KL_divergence}. 
For a study on 1D-correlated datasets, we sampled the distribution associated to random MPS Born machines implemented using the ITensor library~\cite{ITensor_julia}. It is worth noting that the improvement for random MPS is not as significant as other datasets. This can likely be attributed to the lack of significant gaps in the spectrum of the graph Laplacian. 
% We also studied data samples generated from a 1D markov chain to better understand the role of noise in the stability of spectral solutions. We infer that it is generally better to have a significant spectral gap to have a significant and robust improvement. Sec. (\ref{Sec:Stability},\ref{sec: stability_app}) \mr{Not sure this is the right place to discuss this.}.
% We also tested the seriation algorithm on the Bars and Stripes, ground state of the Toric code Hamiltonian~\cite{Kitaev_2003,Verstraete_2006} and the samples coming from a random PEPS~\cite{orus2014practical} architecture. 
A similar behavior can be seen in the data collected from the ground state of random tree-structured Ising Hamiltonians:
\begin{equation}\label{eq:ising_tree_hamiltonian}
    H=-\sum_{i,j}J_{ij}s_{i}s_{j}
\end{equation}
with $s\in\{ -1, 1 \}$, where importantly, not all $J_{ij}$ are non-zero to retain the tree structure. In this dataset, the seriated MPS show significantly lower KL divergence values, as well as vastly improved variances. Lastly, in the case of the BAS dataset, there are large gaps between the low and the intermediate eigenvalues of the graph Laplacian, indicating that if we use the eigenvector corresponding to a smaller eigenvalue for spectral ordering, we are guaranteed to ensure an ordering that preserves locality. This allows allow the seriated model to closely match the BAS dataset given sufficiently large bond dimensions, while randomly ordered models fail to learn the BAS dataset to any meaningful extent for all bond dimensions investigated.
% We also observed similar improvements in the learning of ground state of the Toric code Hamiltonian. These improvements can be qualitatively attributed to the eigenspectrum of the graph Laplacian based on the pairwise mutual information matrix. 
% Lastly, we studied the samples coming from a Ising tree Hamiltonian in a two-fold way. We used the best 1D spectral ordering to demonstrate improvement. We also recovered the tree-like structure where the eigenvectors of the graph Laplacian ensure that the tightly coupled pairs are placed closer than the loosely coupled ones. Moreover, a number of other empirical results also suggest that a multidimensional Laplacian embedding enhances the latent ordering of the data if any. However, there is a need for more careful numerics to further study and develop use cases where the advantage of the multidimensional approach is more significant.
\subsection{Spectral embedding}
\label{sec:spec_embedding}
\begin{figure}[ht]
    \centering
    \includegraphics[width=8.5cm]{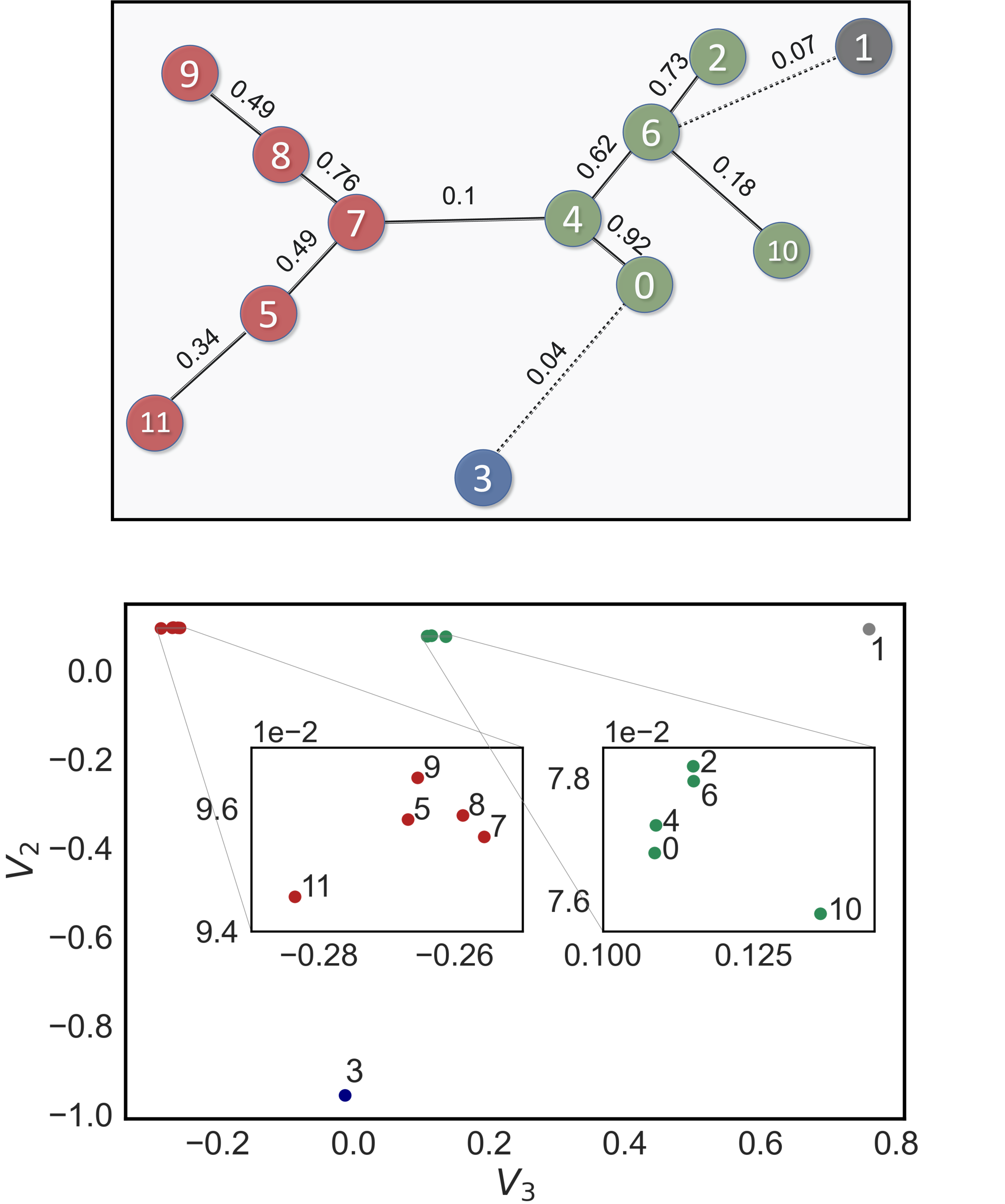}

    \caption{Demonstrating spectral embedding on an Ising tree dataset. (top): The undirected weighted graph is depicted with weights corresponding to the coupling strength $J_{ij}$ of the Ising Hamiltonian. (bottom): Spectral embedding of the sites in the data using the second and the third eigenvectors of the MI graph Laplacian. We use the MI statistics of 1000 bit-strings corresponding to the ground state configurations of the Ising Hamiltonian on $Sz$ basis. Data sites are colored for convenience. The spectral embedding correctly recovers that qubits numbered [7,8,9,5,11] are clustered separately from [2,6,4,0,10] in terms of their relative correlations. It is important to note that closer distance in the above embedding space indicates stronger correlations between the corresponding variables.}
    \label{fig:embedding}
\end{figure}

\begin{figure}[ht]
    \centering
    \includegraphics[width=8cm]{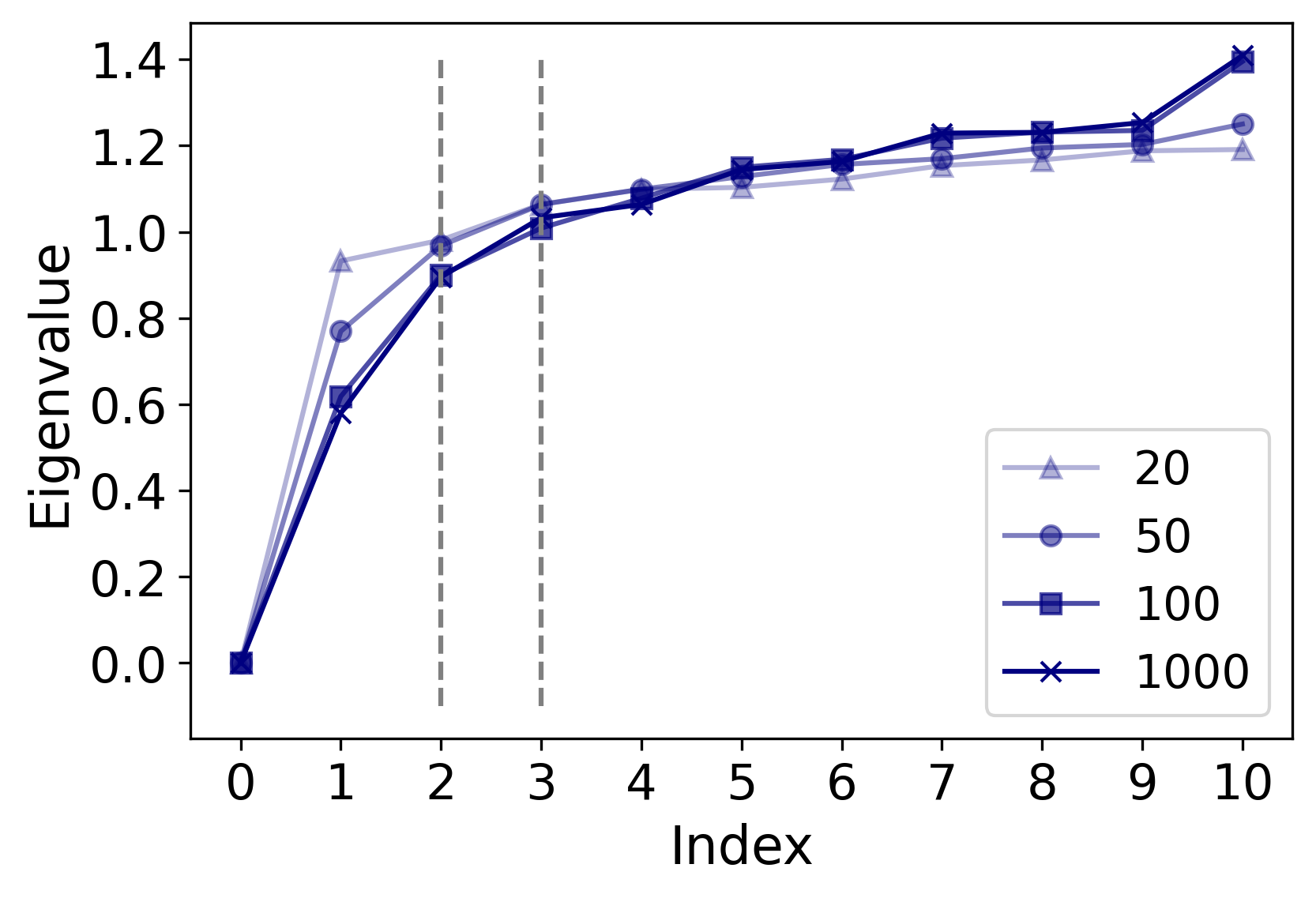}
    \caption{We see the lessening of spectral gap ($\lambda_1-\lambda_2$) of the normalized Laplacian if we construct noisier Laplacians by using fewer samples to estimate the pairwise MI matrix.
    This also means that the Fiedler vector is not a unique vector that will seriate the qubits, meaning that slight variations in the data will generate very different permutations as answers to the seriation problem. The data here is taken from a Markov chain, with the legend showing the number of samples used to estimate the pairwise MI.}.
    \label{fig:noise_spectral}
\end{figure}

While the Fiedler vector $\textbf{x}_1$ encodes the optimal site ordering for 1D correlated data, this may not be optimal for more complicated datasets. In such cases, we can utilize higher eigenvectors of the MI graph Laplacian $L$. The subspace spanned by these eigenvectors can be used to obtain the so-called spectral embedding.
% The site indices are then not only projected onto the Fiedler vector, but also higher eigenvectors $\textbf{x}_j$, with $j>1$.
% We begin by choosing first few eigenvectors of the graph Laplacian.
The utility of higher eigenvectors depends on the magnitude of their respective eigenvalues, where the gaps between eigenvalues typically quickly decrease, leading to unstable representations.
% The MI statistics obtained between pairs of qubits using a finite number of samples, end up revealing some structural information i.e. hierarchical connectivity.

As a concrete example, we construct one instance of the Ising tree dataset, given by the ground state of a Hamiltonian described in Eq.~\eqref{eq:ising_tree_hamiltonian}.
% The Ising Hamiltonian can be written as $H=-\sum_{i,j}J_{ij}s_{i}s_{j}$ where $s\in\{ -1, 1 \}$. Importantly, we define not all $J_{ij}$ to be non-zero and, in particular, they are not all equal. 
The particular tree structure and coefficients $J_{ij}$ can be seen in Fig.~\ref{fig:embedding}. 
% The dataset is constructed from samples coming from the ground state of this Hamiltonian.
% \mr{you had $\lambda_1$ eigenvalue be =0. So you need to count 2 and 3 here. 
% Ans: In the methods, we refer x_0 as the first useless. x1 as fiedler, here we use one more i.e. x2
% }, 
Using the spectral embedding on the first two eigenvectors $\textbf{x}_1$ and $\textbf{x}_2$
, we clearly recover the correlation structure of the ground state which closely follows the magnitude of the $J_{ij}$ terms. We highlight clusters of more closely correlated sites by coloring them in both the Ising tree graph and the spectral embedding. The embedding can now be utilized to find sparse graphs connecting all sites such that the original full pair-wise MI graph is well-reflected. 

%We illustrate this by reconstructing the tree-like structure of a dataset originating from samples of a Ising Tree Hamiltonian ground state. The embedding results in placing more tightly coupled pairs closer than the loosely coupled ones.

\subsection{Stability of the spectral solutions}
\label{Sec:Stability}
% \textit{Low noise and inside perturbation regime:} 
The stability of the spectral ordering by the Fiedler vector can be indicated by the spectral gap $\lambda_2-\lambda_{1}$ of the Laplacian $L(W)$. In general the spectral gap ($\lambda_{k}-\lambda_{k-1}$) is used for stability analysis when the algorithm uses the subspace formed by $k$ eigenvectors. If there is degeneracy in the system, then there might be more than one optimal solutions since the solutions form a degenerate subspace. It does not matter which direction is picked within the subspace or if a more general algorithm is employed. However, if there is no gap between the bands of eigenvalues, then it is indicative that there will be no benefit to the algorithm via seriation.

The spectral ordering solution is stable when the magnitude of unstructured noise is less than the spectral gap~\cite{convex_rel}. \begin{equation*}
||\Delta L ||_{F}\leq (\lambda_k-\lambda_{k-1})/\sqrt{2}
\end{equation*} 

% Here, $||L||_{F}=\sqrt{\sum_{i,j}l^{2}_{i,j}}$ denotes the Frobenius norm of the Laplacian matrix. 
% The proof can be underlined here (MAYBE IN APPENDIX). 
% Diagonalizing $L=O\Lambda O^{T}$, where $\Lambda=diag(\lambda_i)$ and the orthogonal matrix $O$ constitutes eigenvectors. Let us construct $L_{in}=O\Lambda_{in}O^T$ where $"in"$ is short for interpolated. While all the eigenvalues are the same $\lambda_{2}=\lambda_{3}=\frac{1}{2}(\lambda_{2}+\lambda_{3})$. This is how the algorithm can fail, since two different directions (associated with the eigenvectors of the Laplacian) can now minimize the ordering objective function:

% \begin{equation}
%     ||L-L_{in}||^{2}_{F}=||\Lambda-\Lambda_{in}||^{2}_{F}=\frac{1}{2}(\lambda_{3}-\lambda_{2})^{2}.
% \end{equation}  

% The final piece of the puzzle lies in taking an arbitrary matrix $\hat L$  with ordered eigenvalues with $ \hat \lambda_{2}=\hat \lambda_{3}$. 
% $\frac{1}{2}( \lambda_{2}-\lambda_{3})^{2}=\underset{\lambda}{ min}( (\lambda_{2}-\lambda)^2 + (\lambda_{3}-\lambda)^2) \leq ( (\lambda_{2}- \hat \lambda)^2 + (\lambda_{3}- \hat \lambda)^2.$  
% Here, we can use the Hoffman-Wielandt theorem to show $\sum_{i}(\lambda_{i}-\hat \lambda_{i})^{2} \leq ||L-\hat L||^{2}_{F}$ and thus finally we get \begin{equation*}
%     \frac{1}{2}(\lambda_3-\lambda_2)^2 \leq ||L-\hat L||^{2}_{F}.
% \end{equation*}
Note that this bound is given for an unphysical noise form (see Appendix~\ref{s:appendix}). More admissible perturbations that preserve symmetry and non-negativity can be studied but it goes beyond the indication of stability one intends to use from the spectral gap in the methods used in our work. We also demonstrate that the sampling noise in the estimation of MI also leads to the lessening of the spectral gap. Thus further leading to an unstable or less useful solution to seriation (see Fig.~\ref{fig:noise_spectral}). 
% \mr{for ... Discuss what the plot shows.} . 
Further details can be found in~\cite{stability_spectral}, where the authors perform a similar analysis of stability but for closely related spectral clustering problems. We present a proof and interpretation of these results in more detail while also noting the special cases which lead to stronger guarantees on the spectral solution to seriation. %\ref{s:appendix}
\section{Conclusions}
\label{sec:conclusions}

In our work, we apply spectral graph theory methods for qubit seriation in tensor network based generative models. Given a dataset, we first used an estimate of the pairwise MI between variables to construct a similarity graph Laplacian. We showed how a sorting method based on the lowest non-trivial eigenvector, the Fiedler vector, could be used to improve the performance of trained generative models, which in principle may be either 1D quantum or quantum-inspired algorithms. We further identified how higher non-trivial eigenvectors of the data-dependent graph Laplacian can be used to obtain further insights into the underlying dataset. %We also outlined the proof of the distance sensitive ordering and showed convex relaxations of permutation vectors into eigenvectors of the Laplacian of the similarity matrix.  

While we demonstrated improved training performance using MPS, other models can also benefit from this. Concretely, clustering algorithms on the spectral embedding of data can be used to design TN architectures and quantum circuit ansätze for ML tasks that are specific to each given dataset. Given that most negative results are derived using uninformed and generic architectures, we are optimistic that our work can lead to improvements in model-data compatibility. We can also leverage the benefits of seriation in generative modeling tasks by classical ML models such as recurrent neural networks and its variants, using the fact that sequential learning is sensitive to data ordering~\cite{Order_matters}. 

\begin{acknowledgments} 
The authors would like to acknowledge Mohamed Hibat-Allah, Vladimir Vargas-Calderón, and Anirvan Sengupta for their insightful discussions.
\end{acknowledgments}
%end of acknowledgements

% \pagebreak

%\onecolumngrid{}

%\clearpage
%\twocolumngrid{}
% \bibliography{refs/qml,refs/Doc_Biblio,refs/refsAQCall_06202017,refs/new}
% \bibliography{refs/qml,refs/refsAQCall_06202017,refs/new}

\bibliography{refs/quantum-ai}

% \clearpage
% \newpage

%\onecolumngrid{}

\appendix

\label{s:appendix}

\section{Spectral ordering}

\label{s:Proof_ordering} 
We are trying to insure that after the data has been ordered, the adjacent qubits are similar where as the distant qubits are less similar. The ordering can be defined as the index permutation $\pi(1,2,...n)=(\pi(1), \pi(2), ... \pi(n) ) $. The permuted mutual information matrix is $(\pi W\pi^{T})_{ij}=w_{\pi(i),\pi(j)}$.

The cost function can then be written as $C(\pi)= \frac{1}{2}\sum_{i,j}(i-j)^2
w_{\pi(i),\pi(j)}$. Using the permutation substitution i.e. $i \rightarrow \permutationinv{i}$ which results in the replacement $\pi(i) \rightarrow i$, we get the expression 
\begin{equation*}
    C(\pi)=\frac{1}{2}\sum_{i,j}(\permutationinv{i} - \permutationinv{j} )^2 w_{i,j}
\end{equation*}
To further simplify the cost function, one can shift the terms as follows. 
$
(\permutationinv{i} -c- (\permutationinv{j}-c))^2$. Since $\sum_{i}\pi(i)=\frac{n(n+1)}{2}$, we can use $c n=\frac{n(n+1)}{2}  \implies c=\frac{(n+1)}{2}$.  \begin{align*}
    C(\pi)=\frac{n^2}{8}\sum_{i,j}(x_{i}-x_{j} )^2w_{i,j}\\
    x_i=\frac{\permutationinv{i} -(n+1)/2}{n/2} \;\;
    \text{s.t.} \: \sum_{i}x_i=0
\end{align*}

We can finally rescale the terms since it doesn't change anything to our optimal solution i.e. the permutations. After scaling appropriately, we can enforce an additional condition on the scaled $x$ in order for it to satisfy $\mathbf{x}^T\mathbf{x}=\sum_{i}x^2_{i}=1$. Note that, so far we are trying to find the optimal solution for the discrete values of $\mathbf{x}$ and it is in fact a combinatorial optimization problem known to be hard~\cite{Linear_Ding}. The new cost function in terms of the rescaled and shifted variables which we still call $x$ can be written as 
\begin{align*}
    C(\pi)=\frac{1}{2}\sum_{i,j}(x_{i}-x_{j} )^2w_{i,j}\\
    \text{s.t.} \: \sum_{i}x_i=0, \sum_{i}x^2_i=1 
\end{align*}

Where the possibilities of $x_{i}$ are still discrete and come from a scaled version of $\frac{\pi^{-1}(i) -(n+1)/2}{n/2}$. However, we can secretly invoke the quadratic laplacian form by following the steps below.

\begin{align}
    C(\pi)=\frac{1}{2}\sum_{i,j}(x^2_{i}+x^2_{j}-2x_{i}x_{j} )w_{i,j}\\=
    \frac{1}{2}(\sum_{i}x^2_{i}\sum_{j}w_{i,j}-2\sum_{i,j}x_{i}x_{j}w_{i,j}+\sum_{j}x^2_{j}\sum_{i}w_{i,j} )\\=
    \frac{1}{2}(2\sum_{i}x^2_{i}\sum_{j}w_{i,j}-2\sum_{i,j}x_{i}x_{j}w_{i,j})\\
    =
    (\sum_{i}x^2_{i}d_{i}-\sum_{i,j}x_{i}x_{j}w_{i,j})\\=\mathbf{x}^{T}(D-W)\mathbf{x} 
    \label{eq:cost_fun3}
\end{align}
We used the definition of the Degree matrix $D=diag( \{d_{i}\} )$ i.e. $d_{i}=\sum_{j}W_{ij}$ where W can be read as the adjacency matrix leading to the graph laplacian $L=D-W$. Since the problem is still discrete, minimizing this along with the constraint that $\sum_{i}x_{i}$ while maintaining $\sum_{i}x_{i}=0$ is hard. %ADD REFERENCE
However, we can relax this condition by  letting $x_i$ be
continuous and $x_{i}\in [-1,1]$. See~\cite{convex_rel},~\cite{Atkins_1998} for a similar analysis that was intended for applications in DNA sequencing.  We can add a lagrange multiplier for the second condition i.e.  $\mathbf{x}^T\mathbf{x}=1$. The cost function reads as \begin{equation*}
    C(\pi)=\mathbf{x}^{T}L\mathbf{x}-\lambda(\mathbf{x}^T\mathbf{x}-1)
\end{equation*} 
This is identical to writing the Rayleigh quotient for L and x and then minimizing it. $R(L,x)=\frac{\mathbf{x}^{T}L\mathbf{x}}{\mathbf{x}^T\mathbf{x}}$, and the stationary points of this cost function occurs when $\mathbf{x}$ is the eigenvector of the positive semidefinite operator $L$. Hence optimality is when  $L\mathbf{x}=\lambda \mathbf{x}$ and $R(L,x)=\lambda$. The lowest eigenvalue of the laplacian is always $\lambda_{min}=0$ with the eigenvector being the $\mathbf{x}_0=(1,1,...,1)^T$. Since $L$ is positive semi-definite, we have $\lambda_{n-1}\geq\lambda_{n-2}\geq\lambda_{n-3}...\geq \lambda_0$. The next $n-1$ eigenvectors are orthogonal to $\mathbf{x}_0$ which then satisfied the $\sum_{i}x_{i}=0$ condition trivially as it can be read as $\mathbf{x}^{T}\mathbf{x}_0=0$. This helps us in identifying that the $\mathbf{x}_{1}$ minimizes the cost function while maintaining the constraints. And the $n$ components of the second smallest eigenvector of the graph Laplacian will provide us with the permutations needed to order the qubits respecting the condition that nearby qubits will have more similarity~\cite{Ulrike}.

There are alternative convex relaxation techniques which are useful to provide stable solutions with specific datasets. 
It uses  $\mathcal{S}_n$ the set of doubly stochastic matrices, i.e. $\mathcal{S}_n = \{ X \in \mathbb{R}^{n\times n} :  X\geqslant 0, X\mathbf{1}=\mathbf{1}, X^T\mathbf{1}=\mathbf{1} \}$ which is the convex hull of the set of permutation matrices.
We can recover the permuation matrix by imosing orthogonality conditions  $\Pi=\mathcal{S}\cap\mathcal{O}$, i.e.~a matrix is a permutation matrix if and only if it is both doubly stochastic and orthogonal. The fact that $L_A \succeq 0$ means that we can directly write a convex relaxation to the combinatorial problem~\eqref{eqn:disc_cost} by replacing $\mathcal{\Pi}$ with its convex hull $\mathcal{S}_n$, to get
\BEQ\label{eq:relaxedPb1}
\BA{ll}
\mbox{minimize} & g^T X^T L_A X g\\
\mbox{subject to} &  X\mathbf{1}=\mathbf{1},\, X^T\mathbf{1}=\mathbf{1}, \, X \geq 0,
\EA\EEQ
where $g=(1,\ldots,n)$, in the permutation matrix variable $X\in\Pi$. By symmetry, if a vector $X y$ minimizes~\eqref{eq:relaxedPb1}, then the reverse vector also minimizes~\eqref{eq:relaxedPb1}. Since this has a significant negative impact on the quality of the relaxation, the authors~\cite{convex_rel} added the linear constraint $e_1^T X g + 1 \leq e_n^T X g$ to break symmetries, which means that solutions where the first element comes before the last one is always picked.

\section{Stability of spectral solution}
\label{sec: stability_app}
\textit{Exact solution}: If the underlying similarity matrix is a pre-R matrix (named after W.S. Robinson~\cite{robinson_1951} who defined the property of these matrices, then spectral ordering can be used~\cite{Atkins_1998}. A matrix is a R-matrix if it satisfies a simple condition stating that $w_{i,j} \leq w_{i,k}$ for $j <k<i$ and  $w_{i,j} \geq w_{i,k}$ for $i<j<k$. 
The mutual information matrix will be a pre-R matrix iff there exist a permutation $\pi$ such that $\pi W\pi^{T}$ is a R matrix. The coefficients of $W$ decrease as we move away from the diagonal. This basically leads to a guaranteed monotonic Fiedler vector.~\cite{Atkins_1998}

% \mr{rather make this a footnote :D})
% \mr{seems like appendix to me, or at least not here in the stability discussion.}

\textit{Low noise and inside perturbation regime:} We also discuss the stability of solutions where the underlying matrix does not have the above properties. The spectral ordering solution can be shown to be stable when there is a significant spectral gap. Further understanding of this is supported by the following analysis where we begin  by 
diagonalizing $L=O\Lambda O^{T}$, where $\Lambda=diag(\lambda_i)$ and the orthogonal matrix $O$ constitutes eigenvectors. Let us construct $L_{in}=O\Lambda_{in}O^T$ where $``in"$ is short for interpolated. While all the eigenvalues are the same except for $\lambda_{k}=\lambda_{k+1}=\frac{1}{2}(\lambda_{k}+\lambda_{k+1})$, e.g. $\lambda_{2}=\lambda_{3}=\frac{1}{2}(\lambda_{2}+\lambda_{3})$. This is how the algorithm can fail to generate stable solution, since two different directions (associated with the eigenvectors of the Laplacian) can now minimize the ordering objective function: If the system is actually degenerate, then we will have to take the degenerate subspace. It doesn't matter which direction is picked within the subspace or if a more general algorithm is employed. However, if there is no gap between the bands of eigenvalues, then it will in many cases lead to the application not bringing any advantage. Since the cost function Eq.~\eqref{eq:cost_fun2} isn't getting lowered by the spectral solutions. 
\begin{equation}
    \Vert L-L_{in} \Vert^{2}_{F}=\Vert\Lambda-\Lambda_{in}\Vert^{2}_{F}=\frac{1}{2}(\lambda_{3}-\lambda_{2})^{2}.
\end{equation}  
$\Vert..\Vert_{F}=\sqrt{\sum_{i,j}l^{2}_{i,j}}$ denotes the Frobenius norm. 
The final piece of the puzzle lies in taking an arbitrary matrix $\hat L$  with ordered eigenvalues with $ \hat \lambda_{2}=\hat \lambda_{3}$. 
$\frac{1}{2}( \lambda_{2}-\lambda_{3})^{2}=\underset{\lambda}{ min}( (\lambda_{2}-\lambda)^2 + (\lambda_{3}-\lambda)^2) \leq ( (\lambda_{2}- \hat \lambda)^2 + (\lambda_{3}- \hat \lambda)^2.$  
Here, we can use the Hoffman-Wielandt theorem to show $\sum_{i}(\lambda_{i}-\hat \lambda_{i})^{2} \leq \Vert L-\hat L \Vert^{2}_{F}$ and thus finally we get \begin{equation}
    \frac{1}{2}(\lambda_3-\lambda_2)^2 \leq \Vert L-\hat L \Vert^{2}_{F}.
\end{equation}

It is useful to work out the sketch of proof of the Hoffman-Wielandt theorem as it has hints towards using more noise robust algorithms for the ordering problem. 
%% My proof: Get it verified
% Direct proof 

Since symmetric matrices are diagonalizable by orthogonal matrices, we can write $L=O\Lambda O^{T}$ and $\hat L=U \Sigma U^{T}$ with $OO^{T}=UU^{T}=\id $.

Frobenius norm for symmetric matrices simplifies as $\Vert A \Vert^{2}_{F}=\tr(AA^{T})=\tr(A^{2})$. The theorem can be written as the following inequation.
\begin{align}
    \tr( (\Lambda-\Sigma)(\Lambda-\Sigma)^{T} ) \leq \Vert L-\hat L \Vert^{2}_{F}\\
    \tr( (\Lambda-\Sigma)^{2} ) \leq \tr((O\Lambda O^{T}-U \Sigma U^{T})^{2} )
\end{align}

Expanding the square leads to elimination of terms, and we are left to show $\tr(\Lambda O^{T}U \Sigma U^{T}O) \leq \tr(\Lambda \Sigma)$. We can introduce a shorthand $X=O^{T}U$ to show this. The brute force approach is to maximize $ \tr(\Lambda X \Sigma X^{T}) $ with the orthogonality constraint $XX^T= \id $. We can use lagrange multiplier $\Lambda_{l}$ (a matrix here), to write the objective as \begin{equation}
    C=\tr(\Lambda X \Sigma X^{T})-\tr(\Lambda_{l}(XX^T-\id) )
\end{equation}

While $\frac{\partial C}{\partial \Lambda_l} $ yields the orthogonality constraint. Setting $\frac{\partial C}{\partial X}=0$ reveals more. \begin{equation*}
    \frac{\partial C}{\partial X}=2\Lambda X \Sigma-(\Lambda_l+\Lambda^T_l)X=0
\end{equation*}

Since $\tr(\Lambda_l XX^T)=\tr(\Lambda^T_l XX^T)$, we can use $\frac{\partial C}{\partial X}=0$ to find the optimal value of $C$ i.e. $\bar C=\tr(\Lambda X \Sigma X^{T})-\frac{1}{2}\tr( (\Lambda_{l} +\Lambda^T_{l})(XX^T-\id) )=\frac{1}{2}\tr( \Lambda_{l} +\Lambda^T_{l} )$.

\section{Algebraic connectivity}
\label{s:Algebraic connectivity}
Algebraic connectivity has been used to indicate the transmission throughput of a grid quantum network~\cite{Liu_alg}. However, there is also a connection that can be seen between the algebraic connectivity of the graph Laplacian constructed using mutual information as a similarity measure between  distributions lying on qubits with the amount of entanglement present in the quantum state. As a numerical experiment, we show the relationship with algebraic connectivity and the bond dimension in a random matrix product state. As the bond dimension increases, indicating a higher entanglement strength, unsurprisingly, we do get the $\lambda_2$ to increase as well. We compute the spectrum and the second smallest  eigenvector of the normalized laplacian $D^{-\frac{1}{2}}LD^{-\frac{1}{2}}$.

\begin{figure}[ht]
    \centering
    \includegraphics[width=8cm]{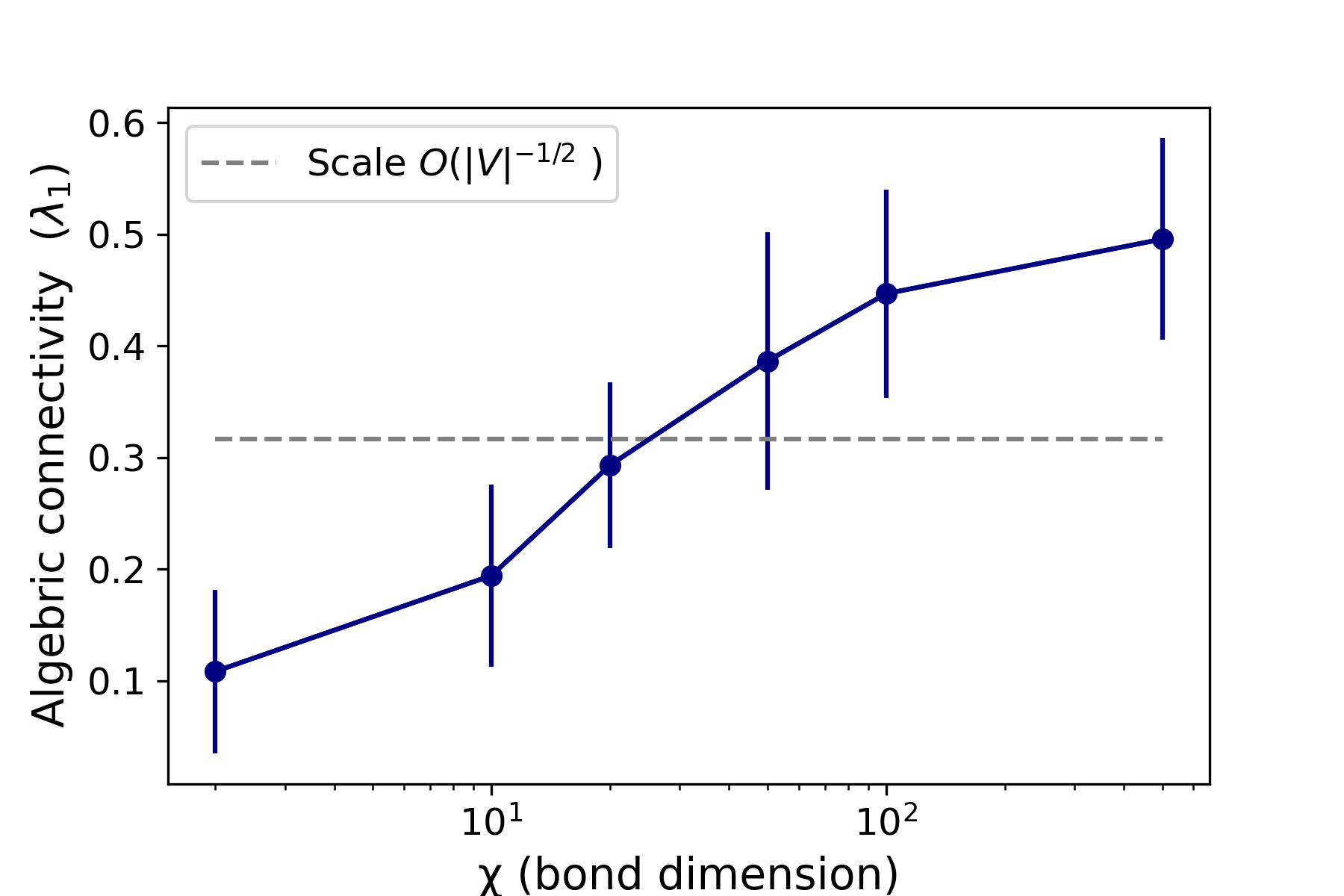}
    \caption{Algebraic connectivity (the first non-zero eigenvalue) against inceasing bond dimension. Here, we take 1000 samples from random MPS with varied bond dimension, estimate pairwise mutual information and then plot the second eigenvalue of the graph laplacian. This indicates that the connectivity increases with increase of entanglement in the quantum state.}
    
    \label{fig:connectivity_bd}
\end{figure}
We can better understand the behaviour of the connectivity of the graphs formed out of 2-point mutual information by monitoring it for varied amount of noise strengths.

% We can use the trace of optimality condition to determine the maximum value.
% $\tr(2\Lambda X \Sigma-(\Lambda_l+\Lambda^T_l)X)=0$. This implies $\tr( (\Sigma \Lambda-\frac{1}{2}(\Lambda_l+\Lambda^T_l) )X)=0$ resulting in 

\end{document}